 \documentclass[smallabstract,smallcaptions]{dccpaper}

\usepackage{url}            % simple URL typesetting
\usepackage{booktabs}       % professional-quality tables
\usepackage{amsfonts}       % blackboard math symbols
\usepackage{nicefrac}       % compact symbols for 1/2, etc.
\usepackage{microtype}      % microtypography
\usepackage{xcolor}         % colors
\usepackage{enumitem}
\usepackage{wrapfig}
\usepackage{subfigure}
\usepackage{algorithmic}
\usepackage{graphicx}
\usepackage{amsmath,bm}
\usepackage{multirow}

\newlength{\figurewidth}
\newlength{\smallfigurewidth}

\begin{document}

\title
{\large
\textbf{Rate Distortion Characteristic Modeling for Neural Image Compression}
}

\author{%
Chuanmin Jia$^{\dag}$, Ziqing Ge$^{\ast,\star}$, Shanshe Wang$^{\dag,\ddagger}$, Siwei Ma$^{\dag,\ddagger,\sharp}$, Wen Gao$^{\dag,\ddagger}$\\[0.5em]
{\small\begin{minipage}{\linewidth}\begin{center}
\begin{tabular}{ccc}
$^{\dag}$School of Computer Science, Peking University, Beijing, China\\
$^{\ast}$University of Chinese Academic of Sciences, Beijing, China\\ 
$^{\star}$Institute of Computing Technology, Chinese Academy of Sciences, Beijing, China\\ 
$^{\ddagger}$Peng Cheng Laboratory, Shenzhen, China\\ 
$^{\sharp}$Information Technology R\&D Innovation Center of Peking University, Shaoxing, China\\
%{\{cmjia, sswang, swma, wgao\}@pku.edu.cn, ziqing.ge@vipl.ict.ac.cn}\\
\thanks{The first two authors contributed equally. Correspondence to: Siwei Ma (swma@pku.edu.cn).}
\small
\end{tabular}
\end{center}\end{minipage}}
}

\maketitle
\thispagestyle{empty}

\begin{abstract}
End-to-end optimized neural image compression (NIC) has obtained superior lossy compression performance recently. In this paper, we consider the problem of rate-distortion (R-D) characteristic analysis and modeling for NIC. We make efforts to formulate the essential mathematical functions to describe the R-D behavior of NIC using deep networks. Thus arbitrary bit-rate points could be elegantly realized by leveraging such model via a single trained network. We propose a plugin-in module to learn the relationship between the target bit-rate and the binary representation for the latent variable of auto-encoder. The proposed scheme resolves the problem of training distinct models to reach different points in the R-D space. Furthermore, we model the rate and distortion characteristic of NIC as a function of the coding parameter $\lambda$ respectively. Our experiments show our proposed method is easy to adopt and realizes state-of-the-art continuous bit-rate coding performance, which implies that our approach would benefit the practical deployment of NIC. 
%In addition, the proposed model could be applied to NIC rate control with limited bit-rate error using a single network.
\end{abstract}

\Section{Introduction}
Lossy image compression locates at the central stage of the visual information processing chain, bridging the upstream computational imaging and downstream visual analytics. Recent years have witnessed an unprecedented booming of end-to-end optimized neural image compression (NIC) approaches using deep neural network (DNN) based auto-encoders with entropy penalized latent variables. Evidence has been shown that the end-to-end data trained convolutional auto-encoder based compression solutions~\cite{balle2017end,balle2018variational,minnen2018joint,balle2020nonlinear,zhao2021learned,cheng2020learned} outperform the classical standardized codecs~\cite{bross2021developments,sullivan2012overview} in terms of rate-distortion (R-D) performances in variety of quality metrics.

% SOTA learned image codecs development
% Tremendous contributions and investigations have been dedicated to NIC research along with the resurge of deep neural networks, ranging from network architecture improvement~\cite{ma2020end,tschannen2018deep,li2018learning,chen2021end}, entropy model approximation~\cite{minnen2018joint,mentzer2018conditional,cheng2020learned}, quantization continuous relaxation~\cite{balle2016end,theis2017lossy,agustsson2017soft,agustsson2020universally,guo2021soft} and rate control strategy~\cite{toderici2015variable,choi2019variable,lee2018context,li2020efficient}. Earlier contributions focus on the framework itself which aims at end-to-end trainable capability~\cite{balle2016end,theis2017lossy,toderici2017full}. The entropy model with advanced bit-rate estimation ability receives dominant attention in subsequent research~\cite{balle2018variational,cheng2020learned,li2018learning} for R-D performance optimization because of the fact that minimizing the relaxed differential entropy with distortion is equivalent to minimizing the upper bound of the actual R-D objective. Novel entropy models usually derive variational upper bound for the differentiable entropy of the latent representations. 

%  major problem
Existing DNN based lossy image compression approaches tend to train multiple models to reach different bit-rate points in R-D space by alternating the hyper-parameters in a pre-defined finite set to control the rate and distortion trade-off for the training objective.
Three categories of drawbacks could be observed given such design. First, the essential analysis for R-D characteristic of NIC behavior is still remaining unexplored, even if pioneer research has been tried to reduce the number of models to realize multiple bit-rate compression~\cite{toderici2015variable,balle2020nonlinear,choi2019variable}. Second, different from conventional standardized codecs that have faithfully provided representative disciplines and methodologies in interpreting the R-D characteristic~\cite{he2001unified,ma2005rate,li2014lambda}, the underlying statistical R-D model of NIC also seems to be less studied in the context of in-depth understanding for the data driven image codecs. Third, the effective rate control model of NIC is missing, which hinders the practical utility of NIC method.
% Contribution
\begin{figure}
\centering
\includegraphics[width=0.7\textwidth]{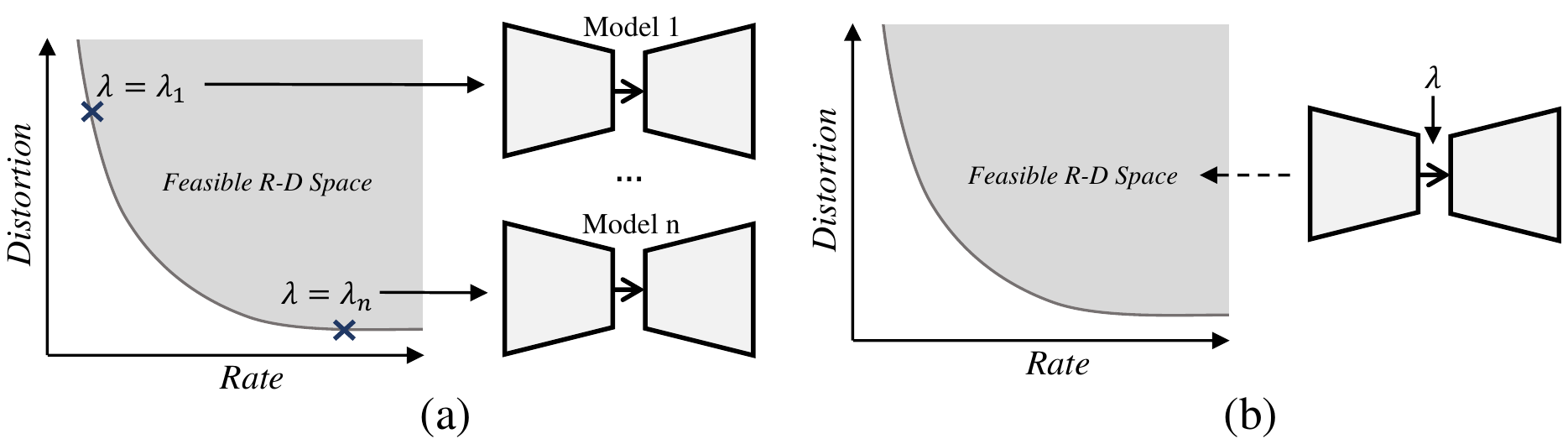}
\caption{Characteristic modeling for NIC in feasible R-D space ($\lambda$ indicates different bit-rate points). (a) Existing approach (multiple models). (b) Our approach (Single model).}
\label{fig:teaser}
\end{figure}

In this paper, we propose a learnable module, Modulation Network (ModNet), to learn a binary representation for the relationship between the bit-rate and the latent variable of auto-encoder. Furthermore, we show the essential R-D behavior of NIC, modeling the rate and distortion characteristic of NIC as a function of the coding parameter $\lambda$ respectively. We also apply the proposed model to rate control scenario for NIC utility. The main contributions of this paper are three folds. {\bf First}, we design and propose \textit{ModNet}, which enables the continuous and arbitrary bit-rate compression of NIC for a wide range of coding bits using a single data trained image codec. The proposed ModNet could be trained individually without any modifications to the pre-trained encoder and decoder. {\bf Second}, we provide the first R-D characteristic view on end-to-end learned image compression. The unexplored rate and distortion behavior models are derived as a function of the coding parameter $\lambda$ respectively, which is informative and discriminative. {\bf Third}, the NIC rate control could be elegantly achieved using the proposed R-D model with limited bit-rate error (BRE), which would definitely benefit the deployment of neural codecs. The proposed method is simple, effective and intuitively appealing, which could be feasibly transplanted to other existing latent variable based compression methods.

\Section{The Proposed Continuous Bit-rate Compression for NIC}
\SubSection{Problem Formulation}
{\bf Preliminaries.} Given the pristine image $\bm{x}$, it is first transformed into latent representation $\bm{y}$ using DNN encoder $\bm{g}_{a}$ and subsequently quantized, yielding discrete-valued latent representation $\hat{\bm{y}}$. Another decoder DNN model $\bm{g}_{s}$, which has reverse architecture with $\bm{g}_{a}$, reconstructs $\hat{\bm{y}}$ to decoded image $\hat{\bm{x}}$. The $\bm{g}_{a}$ and $\bm{g}_{s}$ are jointly optimized using following learning objective in the R-D space:
\begin{equation}
\mathcal{L}(\lambda) = \lambda D(x,\hat{\bm{x}}) + R(\tilde{\bm{y}}),
\label{rd-loss}
\end{equation}
where $\tilde{\bm{y}}$ is the differentiable surrogate entropy of latent representation $\bm{y}$ and $D(x,\hat{x})$ measures the distortion of coded image $\hat{\bm{x}}$. Note that $\hat{\bm{x}}$=$\bm{g}_{s}$($\bm{g}_{a}$($\bm{x}$)) in a sense of auto-encoder. 
%To realize different bit-rate compression, distinct models are required to be trained by varying the $\bm{\lambda}$s in eq.~(\ref{rd-loss}). 
For different bit-rate point (measured by bit-per-pixel, bpp), the $\lambda$ value varies thus different models are needed to be trained. Regarding VBR coding, only one model is used for different bit-rate points.
%In addition, multiple types of distortion metrics could also be deployed to support objective or perceptual based evaluation.

%Regarding the quantized latent representations, it can be losslessly compressed using entropy coding approaches since it is discrete valued. Entropy coding relies on a prior probability model (entropy model) of the quantized latent variables, which is shared at both of the encoder and decoder. However, the quantization operation during training process makes the gradients of the network to be zero almost everywhere, which hinders the back propagation of learning procedure therefore differentiable approximations methods for quantization are mandatory.

{\bf Rate expression.} Entropy estimation of latent representation $\hat{\bm{y}}$ is a challenging but important module in NIC as the entropy model usually follows well-studied prior distribution with statistical closed-form. However, such assumption does not guarantee a perfect match between prior distribution and the distribution of complex content in real world. Denote the image distribution and the distribution of $\hat{\bm{y}}$ as $p_{\bm{x}}$ and $p_{\hat{\bm{y}}}(\hat{\bm{y}})$ respectively. 
%The R-D objective in eq.~(\ref{rd-loss}) is represented by the variational auto-encoder. 
In variational inference context~\cite{balle2018variational}, the posterior distribution $p_{\hat{\bm{y}}|\bm{x}}(\hat{\bm{y}}|\bm{x})$ is approximated using variational density $q_(\hat{\bm{y}}|\bm{x})$  by minimizing the expectation of their Kullback–Leibler (K-L) divergence over the data distribution $p_{\bm{x}}$:
\begin{align}
\mathbb{E}_{\bm{x}\sim p_{x}}D_{KL}(q(\tilde{\bm{y}}|\bm{x})|p(\tilde{\bm{y}}|\bm{x}))=
\mathbb{E}_{\bm{x}_\sim p_{x}}\log p_{\bm{x}}(\bm{x})+ 
\mathbb{E}_{\bm{x}_\sim p_{x}}\mathbb{E}_{\tilde{\bm{y}}_\sim q}[\log q(\tilde{\bm{y}}|\bm{x}) - \notag \\
\log p_{\bm{x}|\tilde{\bm{y}}}(\bm{x}|\tilde{\bm{y}}) - \log p_{\tilde{\bm{y}}}(\tilde{\bm{y}})],
\label{vae-loss}
\end{align}
where $\mathbb{E}_{\bm{x}_\sim p_{x}}\log p(x)$ is a constant value because of the fact that $x$ is determined.
The second term in the KL divergence is equal to zero since $\log q(\tilde{\bm{y}}|\bm{x}) = \log \prod_{i}\mathcal{U}(\tilde{y_{i}}|y_{i}-\frac{1}{2},y_{i}+\frac{1}{2}) = \log 1 = 0$,
% \begin{equation}
% \log q(\tilde{\bm{y}}|\bm{x}) = \log \prod_{i}\mathcal{U}(\tilde{y_{i}}|y_{i}-\frac{1}{2},y_{i}+\frac{1}{2}) = \log 1 = 0,
% \label{uniform-noise-log-q}
% \end{equation}
where $\mathcal{U}$ denotes the additive standard uniform distribution centered on $y_{i}$. The $-\log p_{\tilde{\bm{y}}}(\tilde{\bm{y}})$ indicates the estimated bit-rate of the latent representation of the auto-encoder. Similar to~\cite{balle2018variational}, minimizing the log likelihood of $\bm{x}|\tilde{\bm{y}}$ is identical to minimize the expected distortion of the reconstructed image $\tilde{x}$. As such, using the squared error distortion metric is equal to choosing a Gaussian prior (weighted by $\lambda$) distribution:
\begin{equation}
p_{\bm{x}|\tilde{\bm{y}}}(\bm{x}|\tilde{\bm{y}}, \bm{\theta}_{g})=\mathcal{N}(\bm{x}|\tilde{\bm{x}},(2\lambda)^{-1}\bf{1}),
\label{gaussian-prior}
\end{equation}
where $\tilde{\bm{x}}=\bm{g}_{s}(\tilde{\bm{y}};\bm{\theta}_{g})$ and $\bm{\theta}_{g}$ encapsulates all of the learnable parameters of $\bm{g}_{s}$. 

\SubSection{Our Method}
\begin{figure}[]
	\centering
	\includegraphics[width=0.76\textwidth,height=0.18\textheight]{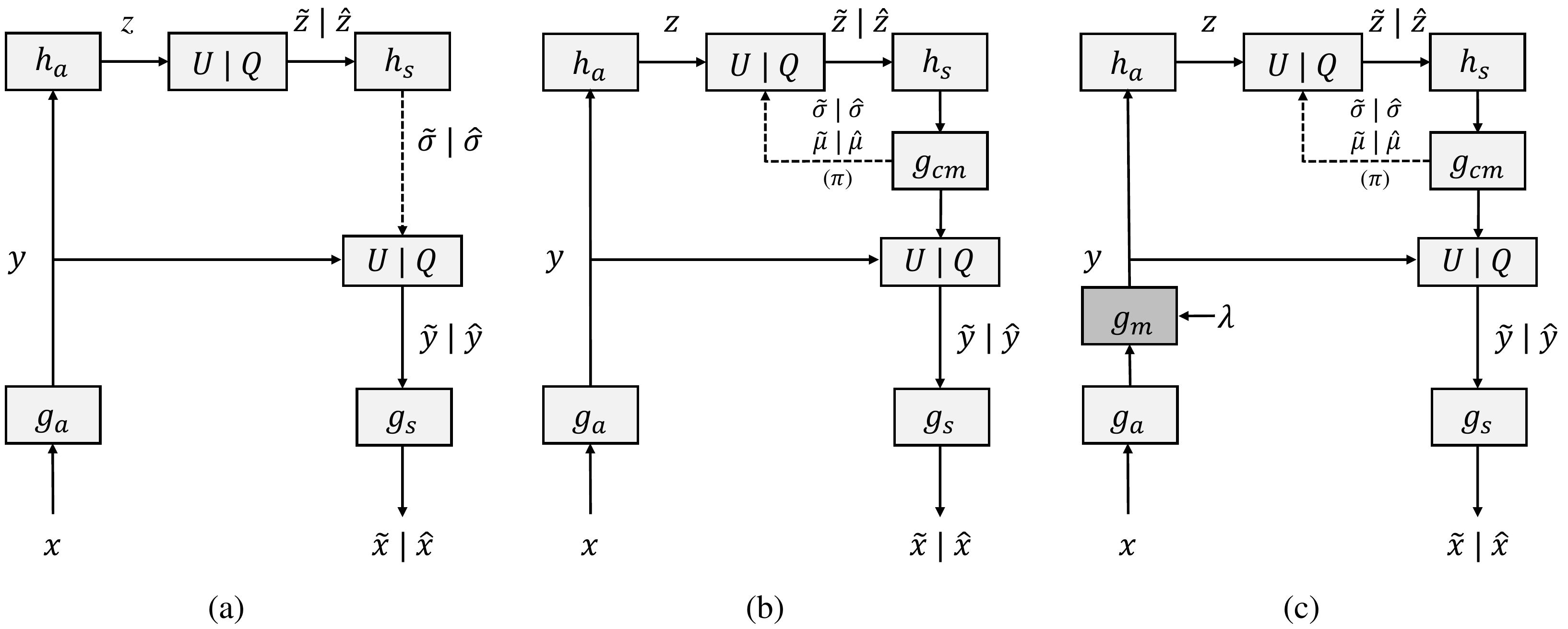}
	\caption{The operation diagram and intuitive comparisons. (a) variational auto-encoder with scaled hyper-prior~\cite{balle2018variational}; (b) joint auto-regressive context model plus hyper-prior~\cite{minnen2018joint,cheng2020learned} ($\pi$ represents the Gaussian weights); (c) Our scheme. The major advantage of our method lies in the ModNet $\bm{g}_{m}$ denoted in darker gray, which is able to support continuous VBR coding using a single trained compressive auto-encoder conditioned on the input $\lambda$ values.}
	\label{fig:infra_compare}
\end{figure}

% from binary representation and rate
It could be easily observed that eq.~(\ref{gaussian-prior}) and eq.~(\ref{rd-loss}) share a common interpretation that the value of $\lambda$ is the deterministic factor for the trade-off between distortion and estimated bit-rate. Thus existing approaches train the learning objective  eq.~(\ref{rd-loss}) using different values of $\lambda$ for different bit-rate points in the R-D space. The purpose of such design is to learn a corresponding set of entropy parameters for a certain degree of distortion, lacking feasibility in VBR coding. However, the cumulative density model has been used in~\cite{balle2018variational} such that it is able to model any univariate density with arbitrary precision. We argue that the constraint of the density has guaranteed the entropy model to be expressive enough to capture a wide range of bit-rate. Because of the fact that density function is supposed to be non-negative and monotonically increasing. The range of the density is also restricted between 0 and 1 with all the Jacobian elements to be non-negative. Therefore, modulating the density model during learning and inference offers the continuous VBR coding ability for entropy model if properly designed.

Different from existing research, we introduce a novel view of eq.~(\ref{gaussian-prior}) that the value of $\lambda$ should not be a fixed-value trade-off factor in the training objective $\mathcal{L}$. Instead, the bit-rate of $\tilde{\bm{y}}$ could be dynamically modulated in both training and inference by external conditions rather than constraint by a hyper-parameter. To realize such objective, we propose to manipulate the elements in the latent representation $\tilde{\bm{y}}$ to adjust the distribution of different elements such that different coded entropy could be obtained by establishing an indication model from $\tilde{\bm{y}}$ to bit-rate.

% map target T to binary representation
Each element $\tilde{y}_{i}$ of $\tilde{\bm{y}}$ is modeled and associated with a binary modulator $\bm{bm}$ before estimating its bit-rate using entropy prior. By using the binary modulator $\bm{b}_{i}$ for each element, $\tilde{y}_{i}\odot\bm{bm}_{i}$ gives the real latent representation. If $\bm{bm}_{i}$ equals to 1, the value at position $\tilde{y}_{i}$ is kept. While $\bm{bm}_{i}$ equals to 0, $\tilde{y}_{i}$ is zeroed. By adjusting the number of zeros in $\tilde{\bm{y}}$, the bit-rate could be dynamic and controllable. 

Another problem to be resolved is to build a model from target bit-rate to the binary representation $\bm{bm}$. Given the knowledge that $\lambda$-model acts as the optimal R-D model in lossy video compression~\cite{li2014lambda}. We proposed to construct a set of trainable parameters together to construct a one-to-one mapping from the coding factor $\lambda$ to $\bm{bm}$. We rephrase the training objective for VBR coding as follows:
\begin{equation}
\mathcal{L}_{vbr}(\lambda) = D(x,\hat{\bm{x}}) + R({\bm{y}}\odot\bm{g}_{m}(\lambda)),
\label{new-rd-loss}
\end{equation}
where $\odot$ represents element-wise product. $\bm{g}_{m}(\lambda)$ indicates the proposed binary modulator which is controlled by coding factor $\lambda$. $\bm{g}_{m}$ is the proposed ModNet (see the right panel of Fig.~\ref{fig:infra_compare}).

\begin{figure}[]
	\centering
	\includegraphics[width=0.7\textwidth]{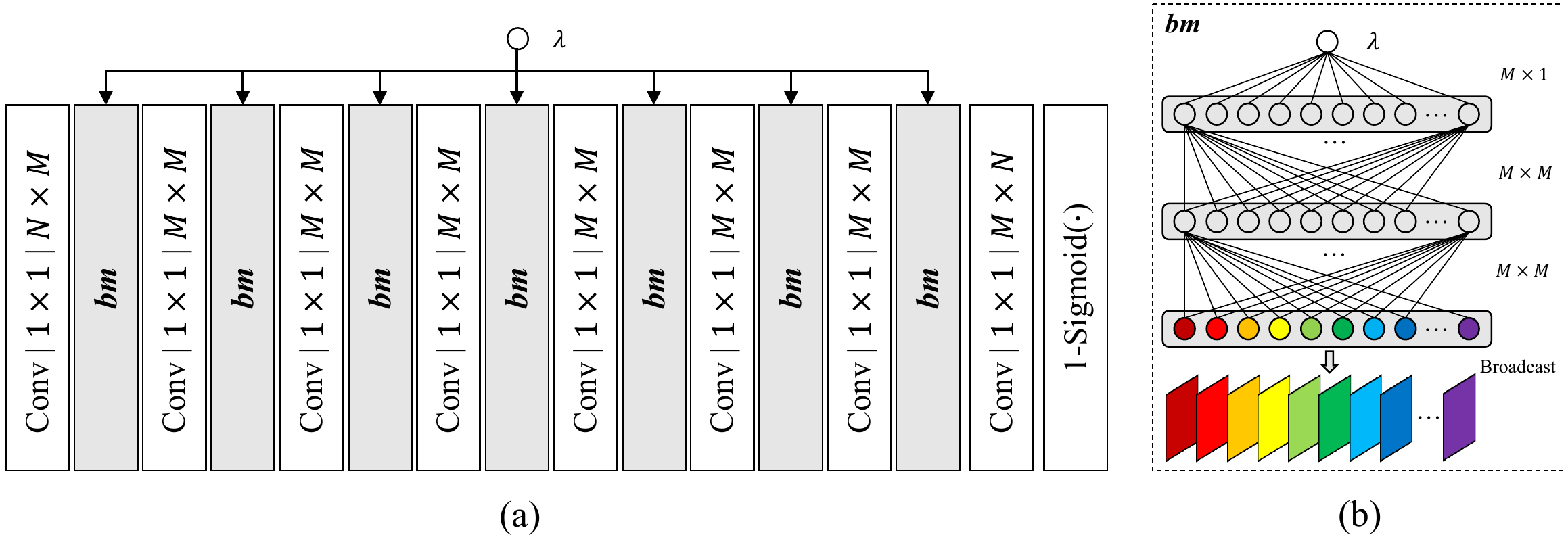}
	\caption{Network architecture of (a) ModNet; (b) binary modulator $\bm{bm}$. ``Conv'' indicates $1\times1$ convolution. $N$ and $M$ denote the number of channels in $\bm{y}$ and the number of intermediate nodes of FCN in $\bm{bm}$ respectively. $\lambda$ is the coding factor.}
	\label{fig:modulation_network}
\end{figure}

% network structure
\SubSection{ModNet}
The proposed ModNet is a plug-in module for the compressive auto-encoders, which consists of two major components, binary modulator $\bm{bm}$ and sequential modulation $\bm{sm}$ for the features, the architecture of which is depicted in Fig.~\ref{fig:modulation_network}. ModNet absorbs the un-quantized latent variable $\bm{y}$ and produces the binarized representation for each position of $\bm{y}$. ModNet has 8 convolution layers separated by 7 $\bm{bm}$ module. Regarding $\bm{bm}$, it is a three-layered fully connected network (FCN), whose input is the single scalar $\lambda$. Each $\bm{bm}$ layer has 100 nodes. The rectified linear unit (ReLU) is chosen for non-linearity except for the last layer where the sigmoid function is used. Therefore the output of $\bm{bm}$ is a 1$\times$100-dim feature vector. Each node of the last layer is then broadcasted spatially. Thus transform the 1$\times$100 vector into a $H_{f} \times W_{f} \times 100$ tensor. The $H_{f}$ and $W_{f}$ indicates the spatial resolution of $\bm{y}$. Similar to $\bm{bm}$, the last layer of ModNet utilizes sigmoid to generate the binary representation for each element of the latent representation.

%The detailed parameter settings of our network are provided in the appendix.

{\bf Binary modulator $\bm{bm}$.} We propose $\bm{bm}$ to transform $\lambda$ into binarized representation for each channel of $\bm{y}$. As such, the arbitrary target bit-rate could be realized by assigning a corresponding $\lambda$ value. The coding factor $\lambda$ is supposed to be a real-value scalar. A dense connected network is designed and proposed to learn the mapping from $\lambda$ into $\bm{bm}$:
\begin{align}
{f}_{1} = \mathrm{ReLU}(\bm{H}^{(1)}\lambda+\bm{b}^{(1)}), \\
{f}_{2} = \mathrm{ReLU}(\bm{H}^{(2)}f_{1}+\bm{b}^{(2)}), \\
\bm{bm} = 1-\mathrm{sigmoid}(\bm{H}^{(3)}f_{2}+\bm{b}^{(3)}), 
\label{modnet}
\end{align}
where ReLU indicates the rectified linear unit activation function. While the $H^{k}$ and $b^{k}$ ($1\leq k \leq 3$) are the affine transform parameters and their corresponding bias. The coding factor $\lambda$ is therefore associated with a tensor $\bm{bm}$. In our design, each spatial channel of $\bm{bm}$ shares the same value for consideration of model capacity and performance.

{\bf Sequential modulation $\bm{sm}$.} The introduction of $\bm{sm}$ is to provide consistent and robust bit-rate modulation performance. It should be particularly noted that simply masking the latent feature $\bm{y}$ using the learned $\bm{bm}$ might result in large a BRE with respect to the target bit-rate. Therefore, we propose to train separate $\bm{bm}$s and cascade them sequentially in a coarse-to-fine manner. As shown in Fig.~\ref{fig:modulation_network} (a), the previous binary modulated features are feeding into subsequent $\bm{bm}$s. Finally, the binary decisions generated by the fully connected layer are broadcasted into the same spatial size as $\bm{y}$. In our design, different $\bm{bm}$s pay attention to different sensitive channels such that bit-rate modulation is comprehensively considered in a channel-wise attention fashion. 
%Further ablations will be provided in experiment sections.

%It is worth noticing that the proposed approach could be harmonized with existing or future entropy models such as the methods in~\cite{balle2018variational,cheng2020learned,minnen2018joint} because ModNet modulates the bit-rate smoothly before quantization and entropy estimation (see Fig.~\ref{fig:infra_compare}(c)).

%In summary, the proposed ModNet is dedicated to resolve the problem in existing NIC that multiple number of models are required to be trained for different bit-rate points. ModNet is able to modulate the elements distribution of the latent representation $\bm{y}$ such that a single trained model could realize continuous VBR coding dynamically through manipulating the coding factor $\lambda$. Consequently, the R-D behavior of our approach could also be investigated.

\Section{The R-D Modeling for NIC}
Given the fact that R-D characteristics are governed by the video contents, it is intractable to give a closed-form of formulation to reflect the inherent R-D behavior. However, with the continuous coding capability enabled by the proposed approach in Section 2, we are able to deploy the trained ModNet to investigate and understand the R-D characteristics of NIC from the statistical modeling perspective. Specifically, the distortion-$\lambda$ (D-$\lambda$) model is properly constructed based on large-scale data regression. And the rate-$\lambda$ (R-$\lambda$) model could be derived accordingly to complete a comprehensive framework of $\lambda$-domain R-D analysis. 

{\bf R-$\lambda$ model.} Given the prior knowledge in video coding that the bit-rate versus $\lambda$ relationship obeys the exponential law~\cite{li2014lambda}, we propose to design the following parametric model:
\begin{equation}
\lambda=\alpha*(\exp(\beta R)-1),
\label{r-lambda-model}
\end{equation}
where $R$ is the bpp value and $\alpha$, $\beta$ are the undetermined coefficients that need to be fitted. We should note that these coefficients might be different among various distortion metrics. We only utilize a single $\exp(\cdot)$ function for model simplicity and the reason for keeping constant 1 in the $\log$ term corresponds to the assumption that if $\lambda \rightarrow$ 0, then $R \rightarrow$ 0. Note that obtaining the exact value of $R$ directly is time-consuming. We propose to resolve this dilemma by transmitting the problem of regressing R-$\lambda$ model to D-$\lambda$ model. The reason for doing so lies in the actual entropy decoding time using auto-regressive model is much longer than network inference time. Thus obtaining the distortion is more realistic.

{\bf D-$\lambda$ model.} As shown in Fig.~\ref{fig:teaser}(a), the $\lambda$ value should be the slope of points on the R-D curve. Such that the following differential equation holds:
\begin{equation}
\lambda = -\frac{dR}{dD}.
\label{lambda-pde}
\end{equation}
Given eq.~(\ref{r-lambda-model}) and eq.~(\ref{lambda-pde}), we could derive the parametric D-$\lambda$ model by solving the above differential equation. Therefore we obtain the following result:
\begin{equation}
D = \frac{1}{\alpha \beta} * \log(1+\frac{\alpha}{\lambda}).
\label{d-lambda-model}
\end{equation}

\begin{figure}[!htb]
    \centering
	\subfigure[]{
	\includegraphics[width=0.3\textwidth]{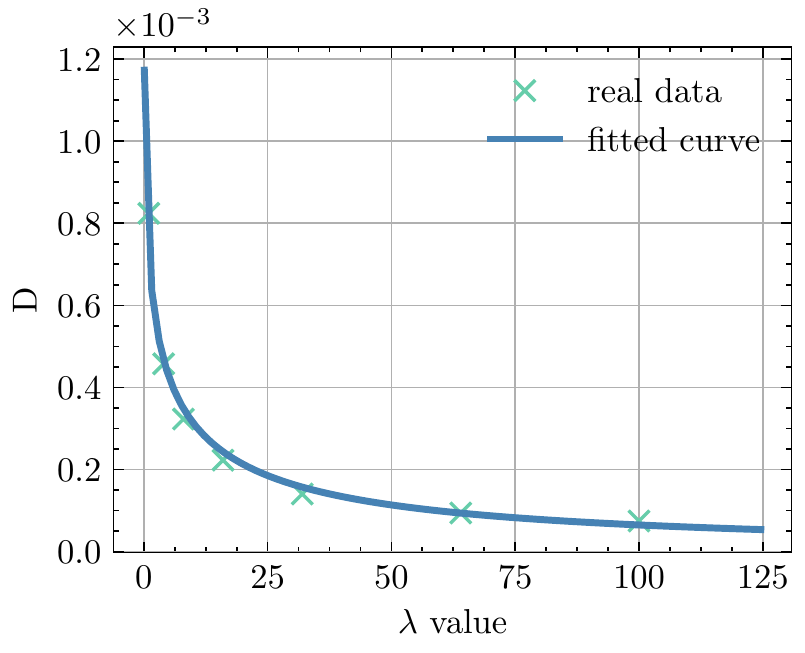}}
	\subfigure[]{
	\includegraphics[width=0.3\textwidth]{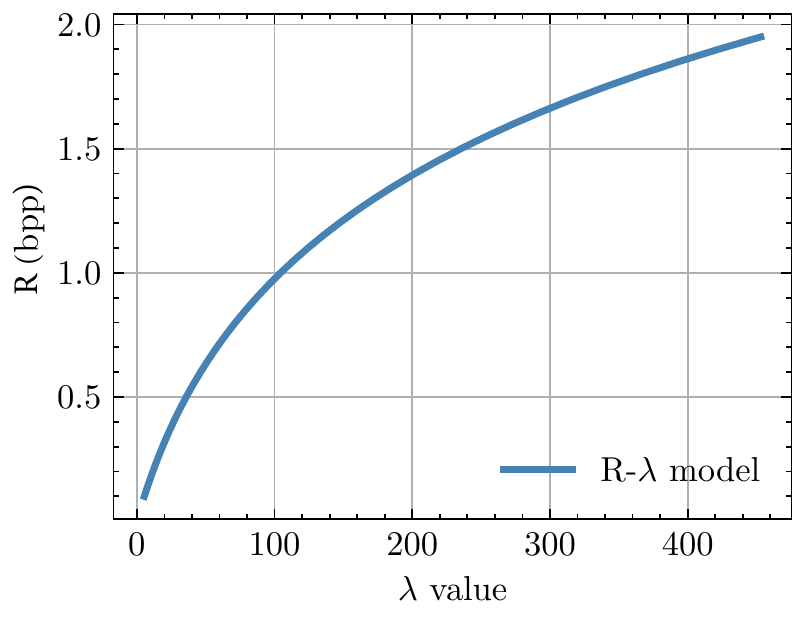}}	
	\caption{R-D behavior of NIC. (a) D-$\lambda$ model; the D-$\lambda$ relationship is fitted. Each cyan cross indicates the averaged data point over all training samples with an identical $\lambda$ value. ($\alpha=39.301$ and $\beta=1.296$ in this MSE based model) (b) R-$\lambda$ model is subsequently derived.}
	\label{r-d-models}
\end{figure}

To obtain $\alpha$ and $\beta$, we employ the numerical method to fit the data points encoded by the proposed continuous method. The typical $\lambda$ values in $\{1,4,8,16,32,64,100\}$ are used to generate the data pairs. For each $\lambda$ value, we compress 600K training samples from~\cite{liu2020comprehensive} using our approach and record the exact value of the corresponding D-$\lambda$ points. The averaged data points for each $\lambda$ are depicted with cyan cross in Fig.~\ref{r-d-models}(a). Note that the detailed descriptions of the dataset and our training method will be elaborated in Section 4. We plot the $D$ (MSE) versus $\lambda$ curve in Fig.~\ref{r-d-models}(a). The $\alpha=89.072$ and $\beta=1.225$ in MS-SSIM. The blue curve represents the fitted model that characterizes the relationship between D and $\lambda$ in NIC, which corresponds to our assumption for the parametric form of R-$\lambda$ model. This phenomenon further indicates that deep codec shares similar characteristics with conventional codec in R-D sense.

\Section{Experiments}
%We conduct experiments by incorporating the proposed ModNet into a representative variational auto-encoder based image codec to obtain continuous VBR coding model. The training details and test results for ModNet are reported. We also provide the fitted coefficients for the R-$\lambda$ model in eq.~(\ref{r-lambda-model}) and D-$\lambda$ model in eq.~(\ref{d-lambda-model}).

%\SubSection{Implementation Details}
{\bf Training method.} We employ the coarse-to-fine training methodology. In particular, the compression model with the highest quality should be obtained and utilized for subsequent training. The representative NIC model~\cite{chen2021end} with the highest bit-rate coding capability (over 1.5 bpp) is firstly trained using loss function eq.~(\ref{rd-loss}). This model is deployed as a pre-trained model for our VBR coding. Note that the proposed ModNet is a plug-in module that aims at modulating the bit-rate of $y$ on condition that the pre-trained model is expressive enough to cover a wide range of bit-rate. As such, the representative model with the highest bit-rate coding capability guarantees this point. The entire network architectures including $\bm{g}_{a}$, $\bm{g}_{s}$, $\bm{h}_{a}$ and $\bm{h}_{s}$ are fixed. Only the weights in ModNet are updated. Subsequently, the ModNet is trained based on eq.~(\ref{training-vbr-loss}). We use different $\lambda$ values within each mini-batch training. The variability of bit-rate is accomplished through different inputs of ModNet:
\begin{equation}
\mathcal{L}_{ModNet} = \Sigma_{i}\mathcal{L}_{vbr}(\lambda_{i}),
\label{training-vbr-loss}
\end{equation}
where $\lambda_{i}$ indicates the sampled $\lambda$ value from a pre-defined $\lambda$-set $\bm{S}=\lbrace n|n\in N^{+},1\leq n\leq 256\rbrace$. To realize un-biased sampling and obtain a relative consistent R-D performance, we employ random sampling approach to select $\lambda_{i}$ from the $\bm{S}$. Note that the $\lambda_{i}$ value chosen for distinct training sample in a mini-batch should not be equivalent. Different mini-batch also has different selected $\lambda_{i}$ values. %Given the proposed approach, the continuous VBR coding could be realized by determining $\lambda$ value before encoding.

\textit{Set-up:} The images in LIU4K dataset~\cite{liu2020comprehensive} are chosen for our training procedure to train the entire network. For training sample preparation, the training images are partitioned into $256\times 256$ patches with no overlap, yielding over 600k image patches.
We set the mini-batch to be 12 and use two NVIDIA GTX 1080Ti GPUs to train the proposed method for 20 epochs. The Adam optimizer is applied for optimization. The initial learning rate is set to be $5\times10^{-5}$ for first the 10 epochs and then halved every 3 epochs. The $N$ and $M$ in Fig.~\ref{fig:modulation_network} are set to be 192 and 100 respectively. 
Different from training a particular model for each bit-rate point, we only train one model for the mean-square-error (MSE) based compression and another one model for Multi-scale Structure Similarity Index Metric (MS-SSIM)~\cite{wang2003multiscale} based compression. Our experiments are based on the RGB domain. The signal-to-noise ratio (PSNR) value is calculated based on restored RGB images. For MS-SSIM, we convert the channel-averaged MS-SSIM value into decibel domain using $-10*\log_{10}$(1-ms\_ssim). Regarding bit-rates, the file size after entropy coding is used to calculate the bpp. %For fixed-rate coding, we train different models for different bit-rate points in both MSE based and MS-SSIM metrics.

{\bf Compression Performances.} Our experiments are evaluated on Kodak image dataset\footnote{http://r0k.us/graphics/kodak/}. In particular, we plot the R-D curves in terms of bpp versus peak PSNR for MSE based training and bpp versus MS-SSIM in terms of MS-SSIM based training. Multiple standardized codecs are compared such as HEVC~\cite{sullivan2012overview} based image codec BPG\footnote{https://bellard.org/bpg/} and VVC Intra~\cite{bross2021developments}. The reference software VTM-8.1 of H.266/VVC is used to reflect the performance of H.266/VVC. The representative learned codec is also compared. Note that we train different models to their convergence for different bit-rate points for compared learned codec and denote it as Fixed-rate coding. For comparisons between fixed-rate coding and the proposed approach, we keep all conditions as same as possible.

\begin{figure}
    \centering
	\subfigure[]{
	\includegraphics[width=0.35\textwidth]{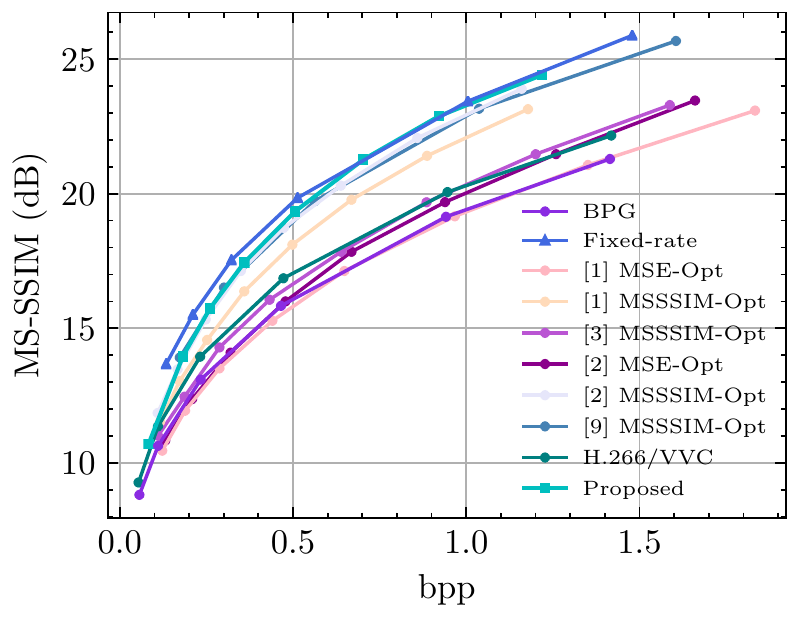}}
	\subfigure[]{
	\includegraphics[width=0.35\textwidth]{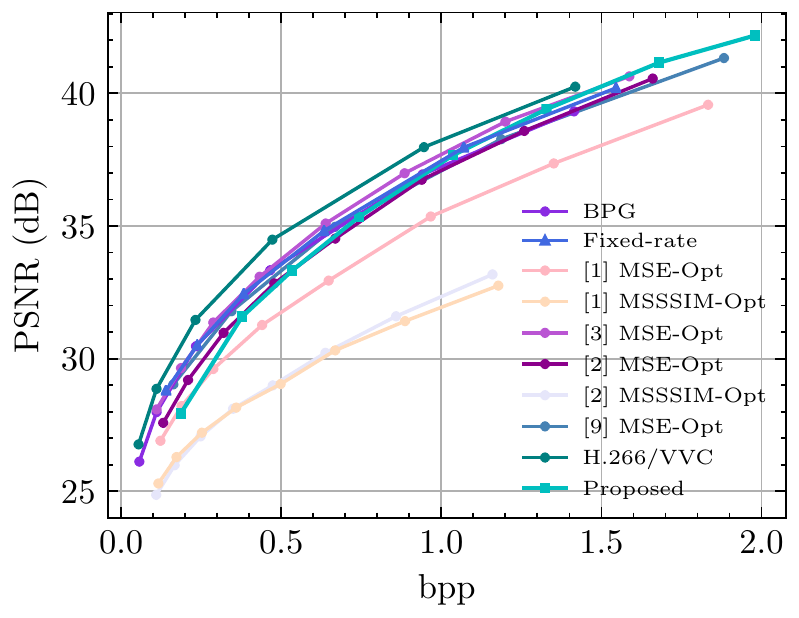}}	
	\caption{Evaluation of the R-D curves on Kodak dataset between the proposed approach and the Fixed-rate coding as well as the conventional standard coding approaches HEVC and VVC using different distortion metrics. (a) MS-SSIM; (b) PSNR. The green, cyan, violet and royal-blue curves indicate VVC, proposed, BPG and fixed-rate coding performance respectively.} %Refer to appendix for more R-D curves comparisons. 
	\label{r-d-curves-comparison}
\end{figure}

Depicted in Fig.~\ref{r-d-curves-comparison}, we plot the R-D curves using MS-SSIM and PSNR metrics respectively. The green curve represents the coding performance of VVC. And the cyan curve reflects the performance of our proposed method. While the violet and royal-blue curves indicate BPG and the fixed-rate coding performance respectively. Regarding MS-SSIM based coding, the proposed method achieves comparable R-D performance against the fixed-rate coding method and outperforms the state-of-the-art method~\cite{choi2019variable}. This phenomenon shows that the proposed method has the ability to cover a wide range of bit-rate compression by using a single trained model. We should also point out that each point within our R-D curve could be realized by modulating the $\lambda$ value of ModNet while the fixed-rate coding needs to train different models to achieve such feature. In addition, both the proposed method and fixed-rate coding significantly outperform the standardized codecs in terms of MS-SSIM, which shows that the neural image codec has better performance on subjective viewing and could be utilized for more practical utilities. Per PSNR based compression, we notice that there still exists a clear margin against VVC intra. The Bjøntegaard-delta bit rate (BD-rate) metric~\cite{bjontegaard2008improvements} performance is provide in Table.~\ref{bd-rates}. The proposed method only results in limited BRE. Moreover, our proposed approach could even realize higher bit-rate using $\lambda$s that are not included in $\lambda$-set, yielding better generalization ability.
\begin{table}[!htb]
\centering
\caption{BD-rate reduction over BPG in terms of MS-SSIM and PSNR metrics}\footnotesize
\begin{tabular}{ccc}
\toprule\hline
           & MS-SSIM  & PSNR     \\ \hline
VVC~\cite{bross2021developments}        & -18.36\% & -19.12\% \\ \hline
Fixed-rate~\cite{chen2021end} & -52.44\% & -0.45\%  \\ \hline
Proposed   & -49.36\% & 1.70\%   \\ 
\hline\bottomrule
\end{tabular}
\label{bd-rates}
\end{table}

\begin{figure}[!htb]
    \centering
	\includegraphics[width=0.55\textwidth]{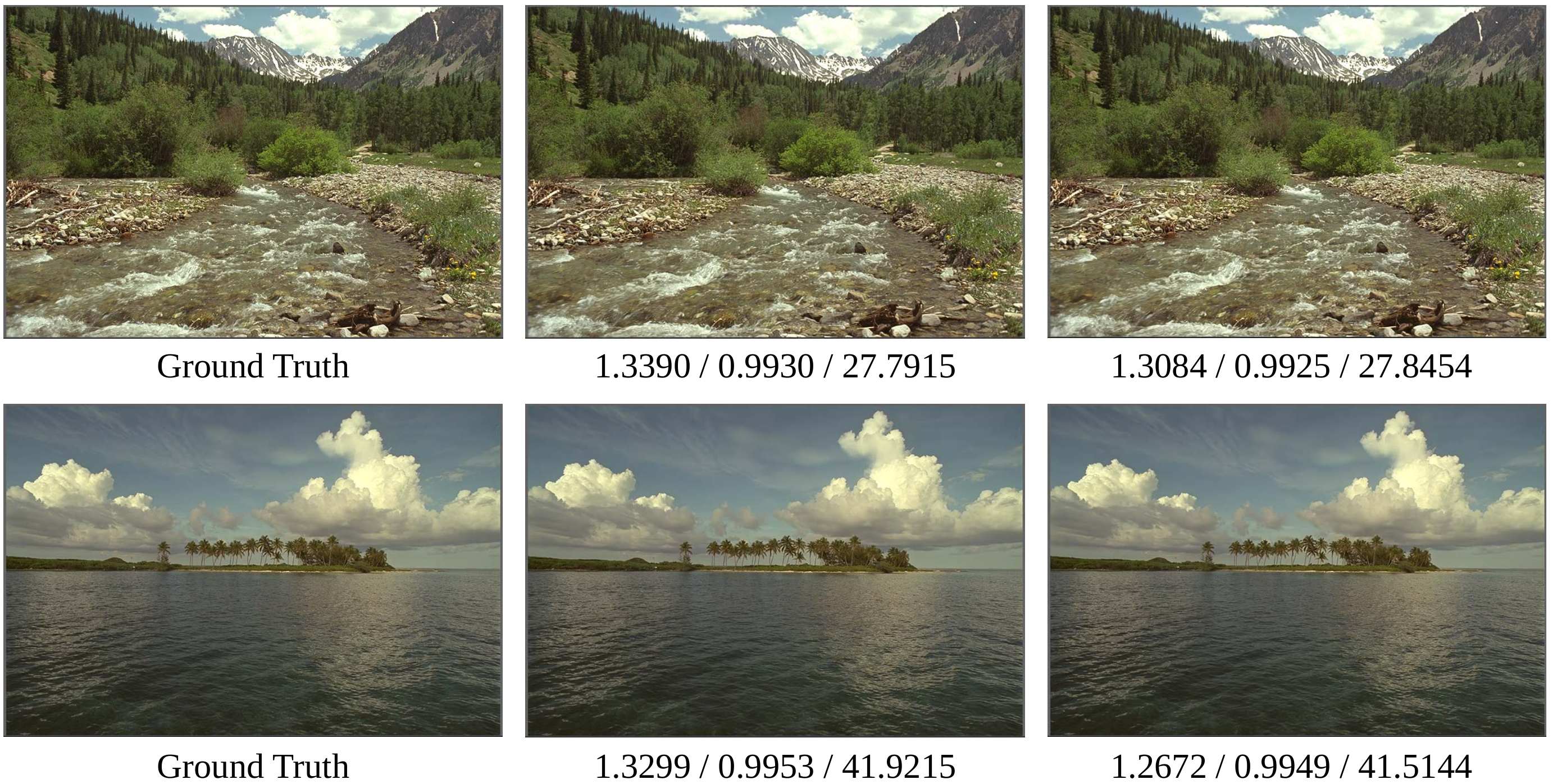}	
	\caption{Qualitative comparisons. The statistics are the values of bpp / MS-SSIM / PSNR (dB). Three panels are ground truth, fixed-rate coding and our methods respectively.}
	\label{visual-comparison}
\end{figure}

{\bf Visualizations.} With the proposed approach, we are able to realize the arbitrary bit-rate coding for NIC. The proposed approach could control the target rate through modulating the $\lambda$ value. To illustrate the effectiveness of our rate control capability in a qualitative comparison manner, we visualize the reconstruction images encoded by fixed-rate models and our method by adjusting $\lambda$ of ModNet to reach a similar bit-rate. As shown in Fig.~\ref{visual-comparison}, we provide two sets of visual comparisons between the proposed approach and the fixed-rate coding, namely Kodim16 and Kodim13 from the Kodak dataset. 
%The first row of Fig.~\ref{visual-comparison} represents the models optimized for MS-SSIM. It is be noticed that both of the objective and subjective quality for the reconstructed images are quite competitive. This shows the superiority of the proposed method. A single continuous VBR coding approach could obtain comparable quality against fixed-rate coding. While the second row of Fig.~\ref{visual-comparison} denotes the decoded images compressed by MSE-optimized models, similar quality can also be perceived. More visual comparisons for low bit-rate coding are provided in appendix.
%Given the direct comparison between fixed-rate and VBR coding of NIC, similar visual quality could be perceived, which shows that the proposed method realizes competitive subjective performance against fixed-rate models.

% \Section{Analysis and Discussion}
% {\bf Advantages.} Based on the architecture comparison in Fig.~\ref{fig:infra_compare}, the proposed method only changes the encoder side of NIC. A modulation module for the encoder is introduced while the decoder structure remains unchanged, therefore a fixed decoder codec is able to decode different bit stream files encoded with different bitrate, without knowing the $\lambda$ used in the encoding procedure. This is the biggest advantage of this VBR methodology because the user client could consequently be very simple. Another advantage of the proposed approach lies in the rate control capability with limited BRE. Such that the number of trained models could be significantly reduced when compared with fixed-rate coding, which would definitely speed-up the progress for employment of NIC.

% {\bf Disadvantages.} Our approach is less efficient against the fixed-rate compression method under low bit-rate coding circumstances, resulting in larger BRE, which is major drawback of our method.

\Section{Conclusion}
For the first time, we investigate the problem of R-D characteristic analysis and modeling for NIC. We propose ModNet for VBR compression via a single network with no modification to the main codec. 
%With such network, we then contribute to make essential mathematical functions to describe the R-D behavior of NIC. 
Moreover, the rate and distortion characteristics of NIC are modeled as a function of the coding parameter $\lambda$ respectively. The proposed method obtains comparable coding performance with fixed-rate coding, which would benefit the deployment of NIC. In addition, the proposed model could be applied to NIC rate control with limited BRE. 
%Regarding future directions, we envision further optimization for the high efficient entropy coding module to reduce the decoding complexity such that NIC would be more suitable for real-world applications.

%\Section{Acknowledgement}\footnotesize
%This work was supported in part by the National Natural Science Foundation of China 62101007 and 61961130392, National Postdoctoral Program for Innovative Talents (BX2021009), and also supported by High-performance Computing Platform of Peking University, which are gratefully acknowledged.

\Section{References}
\bibliographystyle{IEEEbib}\tiny
\bibliography{refs}\tiny

\end{document}